\documentclass[epj]{svjour}
\usepackage{graphics,epstopdf}
\begin{document}
\title{Influence of 
global features of a  Bose-Einstein condensate
on the vortex velocity}
\author{H. M. Cataldo\thanks {\email{cataldo@df.uba.ar}} \and D. M. Jezek\thanks {\email{djezek@df.uba.ar}}}                     
\institute{Departamento de F\'{\i}sica, Facultad de Ciencias Exactas
y Naturales,
Universidad de Buenos Aires, RA-1428 Buenos Aires,  \\ Argentina and 
Consejo Nacional de Investigaciones Cient\'{\i}ficas y T\'ecnicas,
Argentina}
\date{Received: date / Revised version: date}
%
\abstract{
We  study the way in which the geometry of the trapping potential
affects the vortex velocity in a  Bose-Einstein condensate
confined by a toroidal trap.
We  calculate the vortex precession velocity through a simple relationship
between such a velocity and
the gradient of the numerically obtained vortex energy. 
We observe that our results correspond very closely to the velocity
calculated through time evolution simulations. However, 
we find that the
estimates  derived from available velocity field formulas
present appreciable differences.
To resolve such discrepancies, we further study  the induced velocity field,
analyzing the effect of global
 features of the condensate on
such a field and on the precession velocity.
\PACS{
      {03.75.Lm}{Tunneling, Josephson effect, Bose-Einstein condensates
in periodic potentials, solitons, vortices, and topological excitations}   \and
 {03.75.Kk}{Dynamic properties of condensates; collective and hydrodynamic excitations,
superfluid flow}
     } 
} 
\maketitle
\section{Introduction}

Since vortices in Bose-Einstein condensates (BECs)
 were first produced experimentally by Matthews
{\it et al.} \cite{mat99}, many fascinating experiments have been developed 
involving quantized vortices. Most of the first experiments were 
performed in harmonic trapping
potentials, while alternative
trapping potentials have been incorporated  in more  recent experiments.
For example, qua\-dratic-plus-quartic polynomial potentials have been
used to obtain vortex lattices  in fast rotating
condensates \cite{bre04,sto06}.  The quartic  term
has been  introduced to stabilize the system when the angular velocity
exceeds the radial angular frequency of the quadratic term
\cite{fet05,kim05}.
In recent years, there has been a revived interest in observing
vortex states in more complex nonrotating traps,
 due to their relation to persistent currents.
In particular, in recent experiments on toroidal traps, 
 Ryu {\it et al.} \cite{ryu07}
 were the first to observe stable  persistent flow, while
 Weiler {\it et al.} \cite{wei08} have observed the formation
 of vortices.
 
From a theoretical point of view, the dynamics of vortices in  homogeneous 
liquid helium systems
has been widely studied \cite{don91}.
Inhomogeneous superfluids, on the other hand,  may be expected to
display more complex
vortex-dynamics phenomena, since they  give rise to 
several new features
\cite{fet01}.
The experimental achievement of BECs  in different types of 
trapping potentials makes room for a variety of forms of 
inhomogeneous  particle densities. In this context,
the study of the associated vortex velocity field   
in an inhomogeneous  medium
has  acquired increased interest \cite{she041}.
In an axisymmetric trap, a single off-axis vortex exhibits,
 in addition to the ordinary circulating velocity field of a centered vortex \cite{fet01},
 an induced  velocity field \cite{she041}.
In the case of a harmonic trap, such an off-axis vortex 
is subjected to a precession movement \cite{lu00,ge01,an00,fe04,mas06,jeca08}
   related to the above
induced  velocity field, which, evaluated near the vortex position, defines the 
vortex background velocity.
An explicit solution for such an induced field
in  two dimensions,
 within the Thomas-Fermi (TF) approximation,
 was recently published  \cite{she041}. This field  was derived 
from the sole assumption that the divergence of the particle current density vanishes.
We have verified in a previous work \cite{jeca08}
 that such an estimate
 agrees rather well
with the simulation results for a harmonic trap 
near the vortex position, although such results
seem to be partially affected by some kind of boundary effect, disregarded
in the aforementioned theoretical study.
This is more evident for the field far from the vortex, since
a greater correspondence than that arising from  \cite{she041}
was found by considering the field corresponding to
an  antivortex  located
outside the condensate, resembling an image vortex.
Similar studies have been developed
in a recent work by Mason and Berloff  \cite{mason08}
in traps presenting a translational symmetry.

In the present work we  investigate the effect of boundary 
conditions on the vortex velocity
in a toroidal trap.
One would expect that 
for the multiply-connected condensate
formed in that trap, the effect of the boundary conditions
should be much more dramatic than that of previously considered
trapping potentials, 
with consequences on the velocity field all over
the condensate, and
on the vortex precession velocity too. 
Such a
 velocity may be calculated by way of two different approaches \cite{fet01,she041,lu00,ge01}. 
One of them
involves the gradient of the vortex energy and, in the other,
the velocity turns out to be 
proportional to the density gradient. Although the latter method 
appears as more widely utilized,
since it requires the only knowledge of the ground state density,
here we shall show that it may lead to important errors.

This work is organized as follows. In Section~\ref{sec2},
we describe the system and, in particular, the toroidal trapping potential we have utilized.
In Section~\ref{sec3},
we analyze the form of the ground state density depending on the number of particles involved. 
In Section~\ref{sec4}, we present numerical calculations of the vortex precession velocity together
with different approaches to indirect predictions.
In Section~\ref{sec5}, we describe the velocity field, emphasizing the importance of
global effects arising from boundary conditions.
Finally, in Section~\ref{sec6},
we summarize the main conclusions of our study.
 
 \section{Toroidal trapping potential} \label{sec2}
 
We consider  a Bose-Einstein condensate  of Rubidium atoms confined by 
a toroidal trap $V_{\rm{trap}}$.  The
Gross-Pitaevskii (GP) energy
density functional has the
standard form \cite{gros61}
\begin{equation}
  E [\psi]  =\int \left( \frac{ \hbar^2 }{2 m}  |\nabla \psi |^2 +
V_{\rm{trap}} \,|\psi|^2 + \frac{1}{2} g \, |\psi|^4  \right) d^3r ,
\label{ed}
\end{equation}
where $ \psi$ is the condensate wave function and $m$ is the atom
mass. The coupling constant $g $  is written in terms of the
$s$-wave scattering length $a$  as $g=4\pi a\hbar^2/m$, where
$a= 98.98\, a_0 $, which is
the boson scattering length corresponding to $^{87}$Rb 
and $a_0 $ is the Bohr radius.

The variation of $ E $ with respect to $ \psi $, where the number of
particles is fixed, yields the  GP equation \cite{gros61}
\begin{equation}
 \left (-\frac{ \hbar^2 \nabla^2}{2 m}+ V_{\rm{trap}} 
+ g \, |\psi|^2 
 \right ) \, \psi = \mu \psi \; ,
\label{gp}
\end{equation}
where $ \mu $ is the  chemical potential.

We use a toroidal trapping  potential
utilized in  recent experiments \cite{ryu07,wei08},
 which in cylindrical coordinates reads,
\begin{equation}
V_{\rm{trap}}(r,z ) =\frac{ m }{2 } \,\left[\omega_{r}^2 \,  r^2 \,
+ \omega_{z}^2\, z^2\right] + 
V_0 \,\, \exp ( -2 \, r^2/\;\lambda_0^2),
\end{equation}
where $\omega_{r}$  and $\omega_{z}$ denote the radial and axial 
frequencies, respectively. 
We have set $\omega_z >>
\omega_r$ to suppress excitation in the $z$ direction in order 
to be able to use the 2D form of the GP equation \cite{castin}.
As a function of the radial coordinate $r$, 
the 2D potential exhibits a local maximum at 
the center $V_{\rm{trap}}(0) = V_0 $ and an absolute minimum at 
\begin{equation}
r_m =  \frac{ \lambda_0 }{ \sqrt{2} } \, \sqrt{ \ln \left( \alpha \right )},
\end{equation}
with $ \alpha=  4 V_0 / ( m \, \omega_{r}^2 \,  \lambda_0^2) > 1 $.
The corresponding value of the potential is
\begin{equation}
V_{\rm{trap}}(r_m)  =\frac{ V_0 }{ \alpha } \, 
\left[ \ln \left( \alpha \right )  + 1 \right],  
\end{equation}
which verifies $ V_{\rm{trap}}(r_m)< V_0 $, for all $\alpha > 1 $.

 The parameters we have used are as follows:
 $ \omega_r / (2 \pi) =  7.8 $ Hz  and $ \omega_z / (2
\pi) = 173 $ Hz. For the laser beam, we have set 
$ V_0  =  50 $  $\hbar \omega_r$  and $ \lambda_0 = 6 \, l_r $, where
$ l_r = \sqrt{\hbar /( m \omega_r)}$.
Within this set of parameters the minimum of the trapping potential
is located at  $ r_{m} = 5.55 \, l_r $. Hereafter we shall use 
$ l_r$ and $\omega_r^{-1}$ as our length and time units, respectively.
 The minimum value of the potential 
is $ V_{\rm{trap}}(r_m)= 24.4$ $\hbar \omega_r$,
which constitutes a lower bound for the chemical potential.

\section{Ground state}\label{sec3}

\begin{figure}
\includegraphics{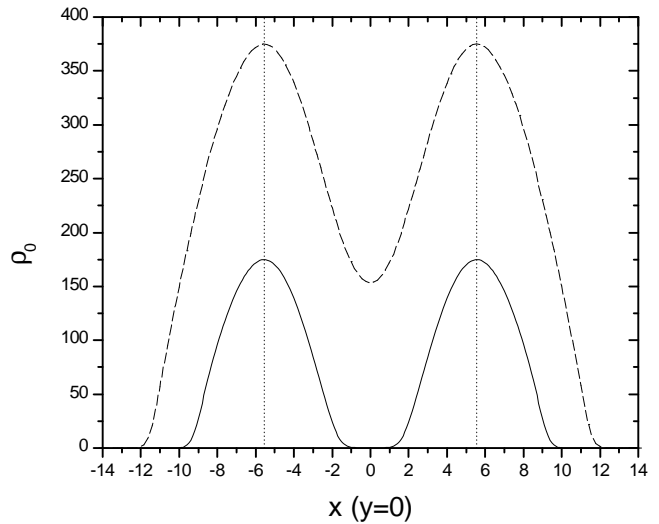}
\caption{\label{fig1b} Ground state density  as a function of the $x$ coordinate
for two different particle numbers.
The full line corresponds to $ N_1 = 3 \times 10^4$ particles, while the dashed line
corresponds to $ N_2 = 10^5 $ particles. We also indicate the positions $ \pm r_m $ of
the trapping potential minimum as vertical dotted lines.
}
\end{figure}
We show in Figure~\ref{fig1b} the ground state density profile
$ \rho_0 = |\psi_0|^2 $ as a function of $x$
at $ y= 0 $ for systems 
formed by two different particle numbers, namely
$N_1 = 3 \times 10^4$   and $ N_2 =  10^5 $.
 These densities have been   obtained by numerically solving
the two-dimensional GP equation, which yields the 
following  chemical potentials:
 $ \mu_1  = 44.5 $ $\hbar \omega_r$ and $ \mu_2  = 67.5 $ $\hbar \omega_r$ 
for the particle numbers $N_1$ and $ N_2 $, respectively.

The form of the ground state density
may be easily analyzed by means of the TF approximation.
In fact, one can obtain the two-dimensional density of the ground state by using the
2D version of  
(\ref{gp}) and  neglecting the kinetic term,
\begin{equation}
 |\psi_0(r)|^2 =  \frac{1}{g_{2D}}\left[ \mu  - V_{\rm{trap}}(r) \, \right]\,
\Theta\left[\mu 
-  V_{\rm{trap}} \right],
\label{TFden}
\end{equation}
where  $g_{2D}= g \sqrt{ m \omega_z / (2 \pi \hbar)} $ \cite{castin}
 and  $ \Theta $ denotes the Heaviside function. Thus, from the above equation we can see that
two types of condensates may arise depending on the value of the chemical potential.
If $ \mu > V_0 $, the condensate is simply connected, while when $  \mu < V_0 $,
the condensate  exhibits a hole around its center. 
In our case, for the number of particles $N_1$ we have 
$  \mu_1 < V_0 $
 and thus the condensate
presents such a hole. On the other hand, for $N_2$ particles, the inequality $  \mu_2 > V_0 $
holds,  and the corresponding condensate turns out to be simply connected.
Finally, we can see from  (\ref{TFden}) that the potential minimum gives rise to
a density maximum. In Figure \ref{fig1b} we have drawn vertical dotted lines
at $ x_m = \pm r_m$, where it may be seen that the maximum density is reached.

\section{ Vortex dynamics} \label{sec4}

In inhomogeneous media,
the vortex precession velocity may be derived by way of two different approaches. One of them
involves the gradient of the vortex energy and the other takes into account 
the background velocity field, along with effects of the vortex core. 
The theoretically derived precession velocity of the latter
method turns out to be 
proportional to the density gradient. In the following Subsections
 we shall analyze  both approaches.

\subsection{ Energy gradient approach}

In inhomogeneous media,
energy $E$ depends on the vortex position. In particular, for a vortex
with vorticity parallel to  the $z$-axis  located at the point  $ \mathbf{r}_0 = (x_0,y_0)$,
 the vortex velocity $ \mathbf{v}_p $ may be 
derived from the following expression  \cite{lu00,ge01}
\begin{equation}
2 \, \pi \, \hbar \, \rho_0 (r_0)\, (\mathbf{\hat{z}} \times \mathbf{v}_p )  =  \mathbf{\nabla}
 E(r_0) 
\;,
\label{vepg}
\end{equation}
where $r_0=|\mathbf{r}_0|$.
An important advantage of this approach is that, for a complex system like the present one,
 this energy can
be calculated numerically. Thus, one can obtain the vortex velocity
 even when  analytic expressions are difficult to derive.
We have numerically computed energy $ E $
as a 
function of the vortex position
for both numbers of particles.
 We have found that  the energy maximum
is located at $r=5.3$ ($r=5.1$) for $ N_1 $ ($ N_2 $).
Such a radius
  turns out to be  slightly shifted with
respect to the density maximum. 
Thus, near such a critical point, the above-mentioned
approaches may give quite different estimates.
\begin{figure}
\resizebox{0.75\columnwidth}{!}{%
\includegraphics{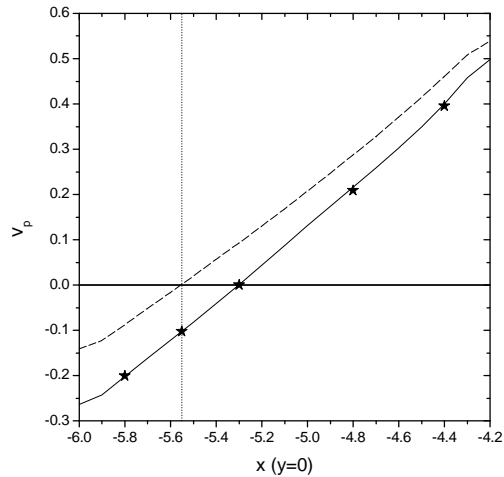}}
\caption{ Vortex precession velocity  as a function
of the $x$ coordinate for the condensate of
$N_1=  3 \times 10^4 $ particles. Values derived from  (\ref{vepg})
are plotted as a full line, while the star dots  indicate the  temporal evolution 
simulations. 
The dashed line corresponds to the estimate given by  (\ref{v1}).
  The position of the density maximum is indicated as a vertical dotted
line.
  }
\label{fig3b}
\end{figure}
\begin{figure}
\resizebox{0.75\columnwidth}{!}{%
\includegraphics{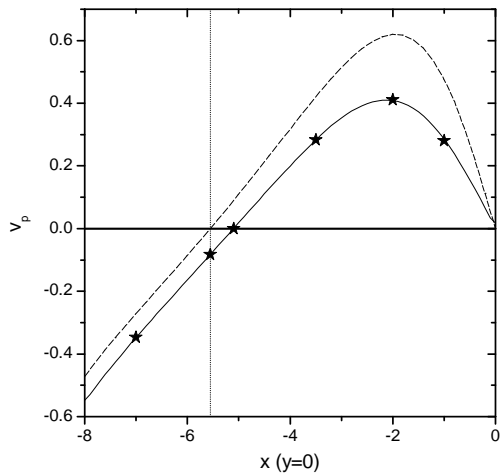}}
\caption{  Same as Figure \ref{fig3b} for
the condensate of $N_2=  10^5 $ particles.   }
\label{fig2b}
\end{figure}
In Figure \ref{fig3b} (Figure \ref{fig2b}), we have plotted the vortex precession velocity 
arising from  (\ref{vepg}) for
the particle number $N_1$ ($N_2$) as a full curve. 
We had to select only a small range
around the density maximum in Figure 2 because the numerical calculation becomes
less accurate at low densities.
We have also
 computed the vortex velocity
for several positions by means of numerical simulations of the
time dependent GP equation. 
Such simulations were performed by means of a fifth-order Runge-Kutta 
algorithm. 
We have considered a system of size 12$\times$12 (15$\times$15),
with a grid of mesh 0.1, for the condensate of $3 \times 10^4 $ ($10^5 $) 
particles.
Identifying the vortex location with the position of the maximum
absolute value of the velocity field, 
we were able to compute the time evolution of
the arc length described by the vortex trajectory.
The precession velocity was then calculated as the
time derivative of the arc length. 
Small fluctuations
on such a velocity ($\sim$ 5\%) were averaged out 
for times much more longer than the typical time-scale of these fluctuations.
Note in Figures \ref{fig3b} and \ref{fig2b}
the 
excellent agreement between such numerical estimates (star dots) 
and the values  predicted by
formula  (\ref{vepg}).
We want to stress that this approach makes 
 no assumptions about velocity fields.

\subsection{ Density gradient approach}

%

Much theoretical work \cite{fet01,she041,lu00,ge01}, based on
different approaches to the velocity field in inhomogeneous media, 
has converged at
the following analytical expression for the vortex precession velocity, which involves
only ground state properties:
\begin{equation}
 \mathbf{v}_{p}   =  \frac{ \hbar   }{ 2 m  }
 \frac{   \mathbf {\hat{z}} \times \nabla \rho_0}
 {  \rho_0  }  \, \,  \ln \left(  \frac{ \xi  \, }
 { R }\right), 
\label{v1}
\end{equation}
where $ \xi $ denotes the local value of the healing length,
\begin{equation}
 \xi =  \sqrt{  \frac{ \hbar^2   }{ 2^{3/2} m \rho_0 g_{2D} }} \, , 
\label{healing}
\end{equation}
and $ R $ the radial size of the condensate. For the present trapping
potential, such a radius 
may be regarded as
the external TF radius, arising from equating to zero expression 
(\ref{TFden}). 
We depict in Figures~\ref{fig3b} and \ref{fig2b} the precession velocity arising 
from equation (\ref{v1}), where
we have used $ R = 9 $ ($ R = 12 $) 
 for the condensate of $N_1$ ($N_2$) particles.
Note that such an expression
vanishes at the density maximum, independently of the value of the condensate
radius.
We may see that this approximation overestimates the precession velocity. 
Here, it 
is important to notice that,
in order to obtain such a theoretical estimate, 
 no boundary conditions
 were taken into account. We believe that this omission may be
 the origin of the above-mentioned discrepancy.
In particular, it is remarkable that the energy maximum, which corresponds to a 
stationary
point, does not coincide with the density
maximum. This suggests that a velocity field due to a global effect, like that 
induced by an image vortex
which would reduce the precession velocity,
may be responsible for this difference. 
 Therefore, we  think that a more careful study of the velocity field is needed.
 In the following
Section we shall develop this idea.

\section{ Velocity field} \label{sec5}

In homogeneous media, a vortex parallel to the $z$-axis  
located at $\mathbf{r}_0$ exhibits the  velocity field 
\begin{equation}
  {\mathbf v}_h(\mathbf{r}) =
\frac{ \hbar   }{  m  } \frac{ 1  } { | \mathbf{r} - \mathbf{r}_0 |^2  } 
\,\, {\mathbf {\hat{z}}} \times
(\mathbf{r} - \mathbf{r}_0).
\label{vfi}
\end{equation}
In trapped gases, the density is no longer homogeneous and an extra velocity field is induced,
which 
we shall call 
the {\em background velocity}  ${\mathbf v}_B(\mathbf{r})$.
If $ {\mathbf v}(\mathbf{r}) $ denotes the total field, then
\begin{equation}
  {\mathbf v}_B(\mathbf{r}) ={\mathbf v}(\mathbf{r}) - 
{\mathbf v}_h(\mathbf{r}).
\label{vf}
\end{equation}
In the following  Subsections we shall distinguish  between two kinds of contributions
to the background velocity,
 one due to local inhomogeneity and the other
related to global effects. 

\subsection{ Local effects } \label{sec5a}

  Sheehy and  Radzihovsky \cite{she041} have 
derived the following approximate expression valid in the TF regime near the vortex core
\begin{equation}
 {\mathbf v}_B(\mathbf{r}) \simeq  \frac{ \hbar   }{ 2 m  }
 \frac{  \mathbf{\hat{z}} \times \mathbf{\nabla} \rho_0(r_0)
 }{  \rho_0(r_0) } \, \ln \left( \frac{  | \mathbf{r} - \mathbf{r}_0  |
 | \mathbf{\nabla} \rho_0(r_0)  |
 }{ 2 \rho_0(r_0)  }\right),
\label{vdd}
\end{equation}
which was obtained through the sole assumption that
 $ \mathbf{\nabla}\cdot \mathbf{j}\equiv 0 $, 
$ \mathbf{j} $ being the particle current density. 
Two observations may be formulated about this expression.
First, the authors have disregarded any boundary condition
 in their calculation,
second, we can see  from the continuity equation
 that the above-mentioned assumption could only be valid
outside the vortex core, where  $\partial \rho_v/ \partial t =  0 $, $ \rho_v$ being the
vortex density.
Therefore,  (\ref{vdd})
may be utilized up to the border of the core at most,
but not inside it, where $\partial \rho_v/ \partial t \ne  0 $. In addition, one should assume 
that global effects are negligible, which, as we shall see, is not the case in the present 
study.
\begin{figure}
\resizebox{0.75\columnwidth}{!}{%
\includegraphics{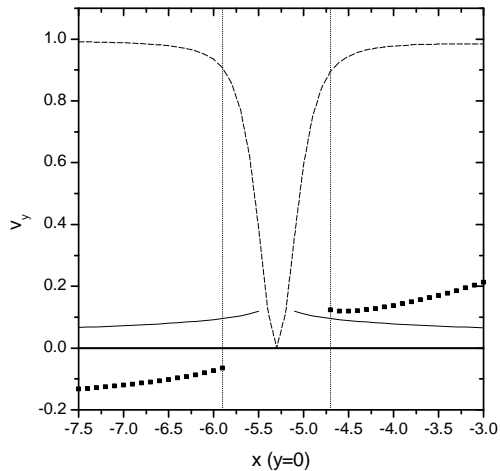}}
\caption{  The $y$-component of the background  velocity
 as a function
of  the $x$ coordinate for a  vortex located at
 $x_0 =-5.3$ and $y_0=0$ in the condensate of $ N_1=  3 \times 10^4$
particles. 
The dots correspond to our numerical results, while the
full line corresponds to  (\ref{vdd}). The function $f$ is depicted by
a dashed line, while the core region can be approximately reduced to the zone 
indicated  within the vertical dotted
lines. }
\label{fig4b}
\end{figure}
In Figure~\ref{fig4b}, we depict the theoretical estimate (\ref{vdd})
together with our numerical results for the stationary point.  
We have not provided any numerical estimate inside the vortex core,
as the wavefunction dramatically
changes 
in that region and the evaluation 
of spatial derivatives involved in the numerical calculation 
of the velocity field is affected by important errors. 
We have also plotted the function $ f = \rho_v / \rho_0 $. 
Note that $f$ is almost equal to unity outside the vortex core, 
which tells us that the densities with 
vortices $ \rho_v $ 
and without  vortices  $ \rho_0 $ turn out to be
almost the same, except in a narrow region inside
the  vortex core, this being a typical situation in the TF regime. 
Thus, the vertical lines in Figure~\ref{fig4b} approximately separate the core
region located between points where $ f \simeq 0.9 $. 
We may observe in Figure~\ref{fig4b} that  the numerically obtained velocity field 
exhibits a  discontinuity around the vortex position, while the theoretical estimate does not.
 In deriving  (\ref{vdd})  the authors have assumed that the ground state density
is a smoothly varying function
 around the vortex core position. 
To further study this issue, we have evaluated the gradient of the
density  $\rho_0$ in the neighbourhood of the border
of the core. We have found that when the $x$ coordinate is varied by $ \pm 0.2$
around $x_0$,
the corresponding velocity arising from  (\ref{vdd}) 
changes about $ \pm 0.1$.
 So, the above discontinuity
could be related to such a gradient, because there is no discontinuity on the 
numerically computed background velocity
along the $y$ coordinate (cf. Figure \ref{fig13b}), which is consistent with the symmetry of the density
at both sides of the vortex core in the $y$ direction.
Note that this local effect correction applies  only around the vortex core, 
while the large
disagreement encountered at larger distances seems to be related to the fact that
the boundary conditions have been disregarded. We shall analyze this issue in the following 
Subsection.

\begin{figure}
\resizebox{0.75\columnwidth}{!}{%
\includegraphics{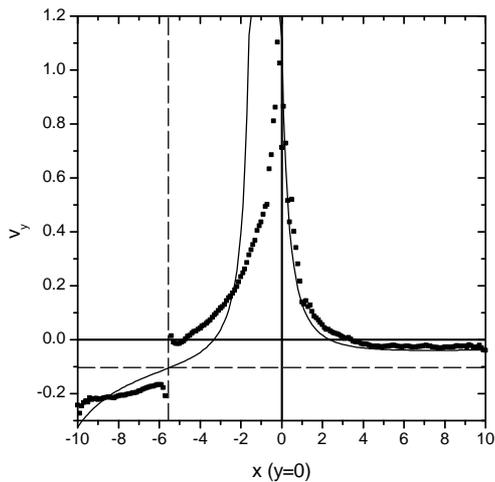}}
\caption{  The $y$-component of the background velocity
as a function
of the $x$ coordinate for
 a  vortex located at $x_0=-5.55$ and $y_0=0$ in the condensate
of $ N_1=  3 \times 10^4$ particles. 
The dots correspond to values arising from numerical simulations, 
while the full line depicts the velocity field of the vortex images,
as explained in the text. 
Dashed horizontal and vertical lines indicate the vortex precession velocity
and the vortex position, respectively.  }
\label{fig10c}
\end{figure}

\begin{figure}
\resizebox{0.75\columnwidth}{!}{%
\includegraphics{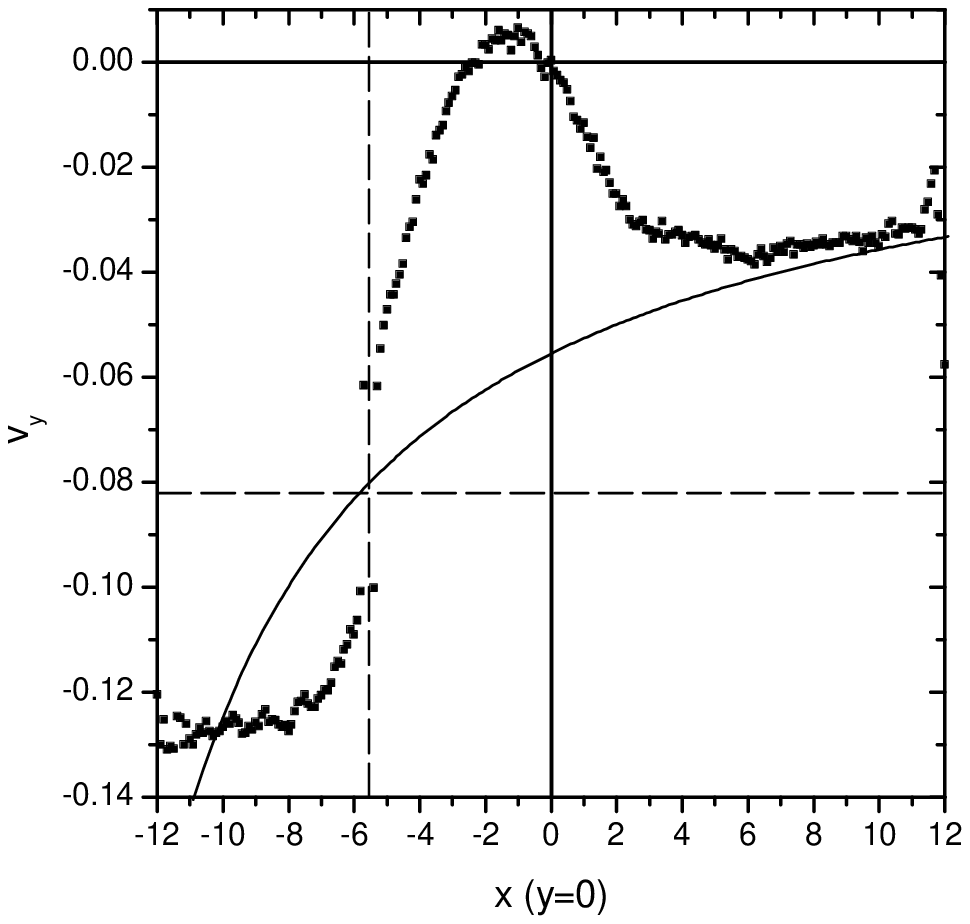}}
\caption{  Same as Figure~\ref{fig10c} for the condensate of
$ N_2=   10^5$ particles.   }
\label{fig12c}
\end{figure}

\begin{figure}
\resizebox{0.75\columnwidth}{!}{%
\includegraphics{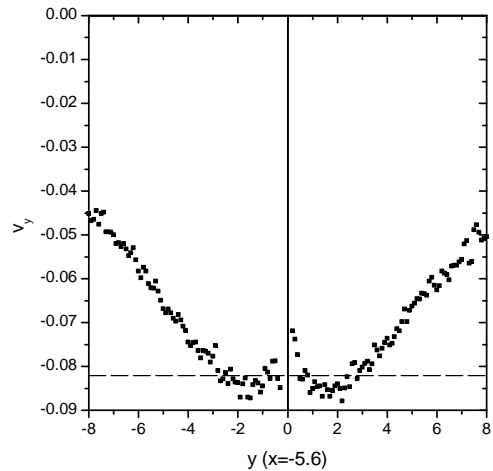}}
\caption{  The $y$-component of the 
background velocity
as a function of the $y$ coordinate for a 
vortex located at $x_0=-5.55$ and $y_0=0$ in the condensate of $ N_2=   10^5$
particles. 
The dots correspond to the results arising from the numerical simulation, while the dashed
 horizontal 
line indicates the precession velocity.   }
\label{fig13b}
\end{figure}

\subsection{ Global effects }

Given the above discrepancy, we believe that a revision of the validity of the
 assumption about the irrelevance of 
boundary conditions is called for. In order to avoid the contribution of local effects, we shall focus 
our attention on points of maximum density,
which cause
  (\ref{vdd})
 to give a vanishing result. 

In Figure \ref{fig10c} we plot the background velocity for a vortex located at the
density maximum
$ \mathbf{r}_0 = (-5.55, 0)$
of the condensate with the number of particles $N_1$. 
First of all, we note that there is again a discontinuity around the vortex position.
Regarding the global shape of the condensate, 
we would like to remind the reader that for this smaller number 
of particles,
the ground state density  exhibits a  hole at its center.
 Thus, this condensate may be roughly approximated by a superfluid 
confined within two coaxial cylinders. 
The boundary conditions for a homogeneous fluid confined within such 
walls involve an infinite number of images. If the inner radius is small enough,
one may consider, as a first approximation, at least three images, namely an external one and two others
located within the inner cylinder, forming a dipole-like image vortex.
If  the  radii of the  internal and external cylinder walls are denoted respectively by
$r_i$ and $ R_e$,  the field produced by each 
image is given by
\begin{equation}
  {\mathbf v}_k(\mathbf{r}) =
\frac{ \hbar q_k  }{  m  } \frac{ 1  } { | \mathbf{r} - \mathbf{r}_k^* |^2  } 
\,\, {\mathbf {\hat{z}}} \times
(\mathbf{r} - \mathbf{r}_k^*),
\label{vim}
\end{equation}
where  their positions $ \mathbf{r}_k^* $ are as follows:
$ \mathbf{r}_1^* = ( -R_e^2 / r_0 , 0 )$, $ \mathbf{r}_2^* = ( -r_i^2 / r_0 , 0 )$,
and $ \mathbf{r}_3^* = ( - (r_i^2 /R_e^2 ) r_0 , 0 )$.
The first image arises from the image of the vortex itself through the cylinder
of radius $R_e$, and thus its charge is $ q_1 =-1 $. The second and the third images
correspond respectively to the
images of the vortex and the previous image
 through the cylinder of radius $r_i$, and thus the
charges are $ q_2 = -1 $ and $ q_3 =  1 $. Therefore, this last pair may be regarded as a
vortex dipole. Note that in order to preserve a fixed value of the circulation of the velocity
field around a closed loop encircling the inner cylinder,  one can only  consider vortex-antivortex
 pairs as image vortices inside such a cylinder. 

In the case of our  condensate, we may roughly estimate the contours through the values 
$R_e=8.5$ and $r_i=2.5$. 
In  Figure \ref{fig10c}, we have also plotted the velocity field of 
the corresponding images as a full line.

  Keeping in mind that the previously reported formula (\ref{vdd}) yields a vanishing
velocity in this case, 
we can see that our theoretical estimate using image vortices plus the above-mentioned gap
certainly provides a  qualitatively  good prediction. 
We also note from Figure~\ref{fig10c}  that the
velocity field due to the images,  evaluated at the vortex position, approximately coincides
with the value of the precession velocity.

Finally, in  Figures \ref{fig12c} and \ref{fig13b}, we depict the $y$ component
of the background velocity as a function of the coordinates $x$ and $y$, respectively,
for a vortex located at the maximum of the ground state density
 of the condensate with $N_2= 10^5$ particles.
In this case the condensate is simply connected, and thus the technique
of using the dipole images, located near $x=0$,  for describing the velocity field
is no longer valid. However 
a `ghost' of such a field may still
 be identified to the right of the vortex in Figure~\ref{fig12c}. 
Although we were not able to model this effect, a single outer image vortex
located at $R_e = 10 $
is shown to provide a good global estimate of the velocity, as seen in Figure~\ref{fig12c}. 
  Moreover, this image provides, again, a good prediction of the
vortex precession velocity marked as a dashed horizontal line in Figure~\ref{fig12c}.
In addition, it may be seen from Figure~\ref{fig13b} that 
such a precession velocity corresponds quite accurately to the
background velocity around the vortex position. 

\section{ Concluding remarks} \label{sec6}

We have seen that 
 the vortex velocity value, stemming from the gradient of the
 numerically computed vortex energy, constitutes a very good estimate of
the vortex precession velocity. We remind the reader that in this calculation, no assumption 
about the vortex velocity field is made.
On the other hand, the expression involving the density gradient,
 in which a model of such a field is perfomed, is shown to
 overestimate the precession velocity.
To resolve this discrepancy, we have investigated 
the influence of boundary conditions on
the vortex velocity field. We have mainly focused on vortices located at
 the maximum of the
ground state  density, 
 where previous theoretical estimates  predicted a vanishing background velocity. 
To roughly describe such a field, we have used image vortices and found 
that several qualitative trends can be succesfully explained,
particularly the lower value of the numerically computed
precession velocity. 
To sum up, we have shown that the vortex velocity field may be affected by 
the condensate boundaries and that such effects can be qualitatively represented by
means of image vortices.

\begin{acknowledgement}
This work has been performed under Grant PIP 5409 from CONICET.
\end{acknowledgement}

\end{document}